\documentclass[journal=jpcafh,layout=twocolumn,manuscript=article]{achemso}

\def\fscale{1}

\usepackage{xcolor}
\usepackage{amsmath}
\usepackage{mhchem}
\usepackage[range-units = single]{siunitx}
\DeclareSIUnit{\debye}{D}
\DeclareSIUnit{\dec}{dec}
\usepackage{hyperref}
\usepackage[english]{babel}
\usepackage{amssymb}

\newcommand{\alq}{\ensuremath{\text{Alq}_\text{3}}}
\newcommand{\irppyacac}{\ensuremath{\text{Ir(ppy)}_\text{2}\text{(acac)}}}

\newcommand{\npb}{\ensuremath{\text{NPB}}}

\title{Enhancement of Spontaneous Orientation Polarization in Organic Semiconductor Mixtures}

\author{Alexander Hofmann}
\author{Albin Cakaj}
\author{Lea Kolb}
\affiliation{Universität Augsburg, Institut für Physik, 86159 Augsburg, Germany}
\author{Yutaka Noguchi}
\affiliation{School of Science \& Technology, Meiji University, Kawasaki, Japan}
\author{Wolfgang Brütting}
\email{wolfgang.bruetting@physik.uni-augsburg.de}
\affiliation{Universität Augsburg, Institut für Physik, 86159 Augsburg, Germany}


\begin{document}

\maketitle

\begin{abstract}

The alignment of permanent dipole moments and the resulting spontaneous orientation polarization (SOP) is commonly observed in evaporated neat films of polar organic molecules and leads to a so-called giant surface potential.
In case of mixed films, often enhanced molecular orientation is observed, i.e.\ a higher degree of alignment, in comparison to neat layers, if it is diluted into a suitable (non-polar) host.
So far, different possible influences on molecular orientation have been discussed, the most prominent probably being the so-called surface equilibration model.
In this contribution, we discuss how surface equilibration can influence orientation in mixed layers, and which other intermolecular interactions have to be considered to explain the observed enhancement of SOP in mixed layers.

\end{abstract}

\section{Introduction}

Organic semiconductors are molecular materials with distinct size, shape and electronic properties, depending on the purpose they were made for.
Of particular interest for application in state-of-the art organic light emitting diodes (OLEDs) are low-molecular weight materials with masses up to about \SI{1000}{amu}\cite{ForrestBook2020}.
Such materials can be processed by physical vapor deposition (PVD), i.e.\ by heating the solid material in (ultra-)high vacuum to evaporate or sublime them such that a beam of molecules is directed toward a substrate being kept at a temperature $T_\mathrm{S}$, on which the molecules condense and form a thin film.
Only in special cases using single-crystalline substrates in combination with highly symmetric molecules, like pentacene or perylene derivatives, ordered {(poly\nobreakdash-)} crystalline films can be achieved\cite{Schreiber_2004}.
The vast majority of molecules, however, grows in a non-crystalline, disordered fashion -- in particular if technical substrates like Si wafers with an oxide layer, glass substrates or transparent conducting oxides on glass are used.
Nevertheless, in many cases, the obtained films are not fully amorphous because the molecules can adopt some preferential alignment along the surface normal of the film such that they appear to be "anisotropic molecular glasses", a term coined by the Ediger group several years ago\cite{Ediger2019}.
They have specifically observed that anisotropy depends in a characteristic manner on the ratio of $T_\mathrm{S}/T_\mathrm{g}$, with $T_\mathrm{g}$ being the glass transition temperature of the bulk material.

Such uniaxial alignment of the molecules in a thin film implies that some of their microscopic properties, like electronic polarizability, (optical) transition dipole moment (TDM) or permanent dipole moment (PDM), exhibit macroscopic anisotropy that can be probed experimentally\cite{Frischeisen2010,Ito2002,Yokoyama2011}.
Note, however, that the alignment of molecules is usually not uniform, i.e.\ pointing all along the same direction, but follows an orientation distribution which is a priori not known.
Thus, one has to combine different measured observables and/or molecular simulations to get deeper insights into the microscopic film structure, as discussed e.g.\ in Reference~\citenum{Hofmann2021}.
Apart from optical anisotropy ("birefringence") \cite{Yokoyama2011} and TDM alignment \cite{Schmidt2017}, orientation of molecules with a non-vanishing PDM can lead to a phenomenon known as spontaneous orientation polarization (SOP)\cite{Noguchi2019}.

Even though the first observation of SOP -- and the concomitant interfacial charges in bilayer structures or the so-called giant surface potential (GSP) on the upper side of a film exposed to vacuum -- dates back more than 20 years\cite{Berleb2000,Ito2002}, the understanding of the driving forces behind it and the engineering of its magnitude have just recently gained increasing attention\cite{Noguchi2020,He2022,Pakhomenko2023}.
Furthermore, the orientation of molecules with PDMs has a huge impact on the electronic devices built using organic semiconductors, which has been shown for OLEDs\cite{Noguchi2013,Bangsund2020} and also in vibration energy generators\cite{Tanaka2020} in many ways.

In this work, we study mixtures of two organic semiconductors, one being polar and the other non-polar, and investigate the resulting SOP of such mixtures in dependence on the concentration of the polar species as well as the film deposition conditions, specifically the substrate temperature $T_\mathrm{S}$ and the \textit{effective} glass transition temperature of the mixture, $T_\mathrm{g,mix}$.
We show that an enhancement of the degree of PDM alignment is possible by co-deposition of the polar molecules with a non-polar host matrix -- an approach termed "dipolar doping" in an earlier work\cite{Jaeger2016}, however, its magnitude seems to depend on the initial degree of alignment in the neat film.
Furthermore, unlike in non-polar emitter molecules studied before by us\cite{Nguyen2023}, the data for samples prepared by variation of the substrate temperature $T_\mathrm{S}$ and those from a concentration variation in higher or lower $T_\mathrm{g}$ hosts cannot be mapped onto each other.
This indicates that in addition to the surface equilibration model put forward by the Ediger group\cite{Dalal2015,Bishop2019}, the description of SOP formation and control requires considering electrostatic dipole-dipole interactions between the molecules' PDMs as well.

\section{Materials and Methods}

\begin{figure}[tbp]
	\centering
	\includegraphics[width=\fscale\linewidth]{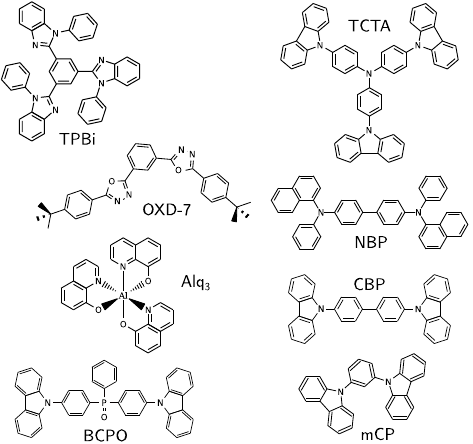}
	\caption{Molecules investigated in the scope of this work. Left side are the polar species TPBi, OXD-7, \alq\ and BCPO (high to low PDM). The conformer shown for OXD-7 is the probably predominant one\cite{Emelyanova2013}. On the right hand side, non-polar host materials TCTA, NPB, CBP and mCP are given, ordered top to bottom from high to low glass transition. For a complete list of parameters please refer to Table~S1 in the supporting information.}
	\label{fig:structures}
\end{figure}

We use TPBi as prototypical polar SOP material and mix it with non-polar carbazole-based host materials, such as TCTA, CBP, NPB and mCP (chemical structures are shown in Fig.~\ref{fig:structures}).
Further polar SOP materials (OXD-7, BCPO, \alq{}) will be used for comparison as well.
TPBi is well known for its GSP in evaporated thin films with a magnitude of about \SIrange{60}{100}{\milli\volt\per\nano\meter}, depending on the details of film growth conditions (mostly $T_\mathrm{S}$ and deposition rate)\cite{Wang2022,He2022}.
Please note that the GSP corresponds to electric fields of the order of \SI{e8}{\volt\per\meter}, which is comparable to fields in OLEDs under operation.

It has recently been shown that TPBi comes in two different isomers\cite{Wang2022}.
One has C3 symmetry, where all three benzimidazole units point to the same half space (i.e.\ all up or down), such that the PDM vector, with a magnitude of \SI{7.6}{D}, is aligned along the C3 symmetry axis and, thus, normal to the central phenyl ring.
The other one has no (or just the trivial C1) symmetry with one of these benzimidazole units pointing to the opposite side so that the PDM vector (with a magnitude of \SI{6}{D}) is not aligned along the C1 axis, which would again be perpendicular to the central phenyl ring.
Due to their slight energy difference, the low-symmetry C1 isomer is much more abundant (\SI{83}{\%}) in evaporated films, which causes a relatively small degree of PDM alignment of about \SI{8}{\%} only (see below for details) in neat films of TPBi.

We study SOP by measuring the surface potential of evaporated thin films using a vibrating Kelvin probe (KP)\cite{Ito2002}.
Therein, PVD of the organic material is performed at variable substrate temperature $T_\mathrm{S}$ between \SI{+80}{\celsius} and \SI{-70}{\celsius} and the contact potential difference (CPD) of successively thicker growing films is measured by the KP system.
Typically, a Si wafer with native oxide layer is used as substrate and the film thickness is increased in steps of \SIrange{10}{20}{\nano\meter} between individual CPD recordings while growth rate is kept constant around \SI{0.3}{\angstrom\per\second}.
It is important to note that the KP system is directly attached to the PVD growth chamber so that there is no vacuum break between these subsequent growth and measurement steps.
Moreover, throughout the whole growth and measurement procedure, the sample is kept in dark or only illuminated by long-wavelength light that is not absorbed by the organic material.

From the (ideally) linear slope of the CPD vs.\ film thickness one obtains the so-called giant surface potential slope, which is proportional to the PDM magnitude according to:
\begin{equation}
m_\mathrm{GSP} = p\left\langle\cos\theta_\mathrm{PDM}\right\rangle{}n / \varepsilon~,
\end{equation}
were $\theta_\mathrm{PDM}$ is the angle of the PDM vector relative to the substrate normal and $\left\langle \dots \right\rangle$ indicates the average over all molecules in a given film volume.
Further, $n$ is the number density of molecules in the film and $\varepsilon$ its dielectric constant $\varepsilon_\mathrm{r}$ times the vacuum permittivity $\varepsilon_\mathrm{0}$.

The equivalent quantity related to the concomitant surface or interface charge density is given by
\begin{equation}
\sigma = m_\mathrm{GSP} \varepsilon~,
\end{equation}
with a unit of \si{\milli\coulomb\per\square\meter}.

As mentioned above, from the known PDM magnitude of the polar molecule (or its average, if there is more than one isomer or different conformers), one can obtain a dimension-less figure of merit for PDM alignment in respect to the substrate normal as:
\begin{equation}\label{eq:orientation_maxgsp}
\Lambda=\left\langle\cos\theta_\mathrm{PDM}\right\rangle=\frac{m_\mathrm{GSP}}{m_\mathrm{GSP,max}}~,
\end{equation}
which can be interpreted as the ratio of actually measured GSP (slope) to the theoretical maximum value for perfectly parallel alignment of all PDM vectors along the substrate normal giving
\begin{equation}
m_\mathrm{GSP,max} = p\rho\frac{M}{\mathrm{N_A}\varepsilon}~,
\end{equation}
with $M$ being the molecule's molar mass, $\rho$ the film density and $\mathrm{N_A}$ Avogadro's constant.
Per convention, $\Lambda$ is positive if the PDMs point away from the substrate, while the unit is often given in \%.
We note that another way to control molecular alignment is by varying the deposition rate, as will be discussed further below.
For example, He~et~al.\ prepared TPBi films using rates from \SIrange{0.05}{10}{\angstrom\per\second} and found out that rate variation is equivalent to a change of $T_\mathrm{S}$ by about \SI{13}{\kelvin\per\dec}\cite{He2022}.
This, in turn, means that we effectively span more than $10$ orders of magnitude in rate by varying the substrate temperature, which is hardly possible to achieve directly.

It is important to recall that PVD grown films of such polar organic molecules are disordered such that the orientation of molecules (and, thus, their PDMs) is never uniform.
Rather, it has to be described by some orientation distribution function $f(\cos\theta_\mathrm{PDM})$, which is a priori unknown.
This means that it is hardly possible to conclude on the shape of $f(\cos\theta_\mathrm{PDM})$, if only the PDM alignment parameter $\Lambda$ is known.
For example, $\Lambda = 0$ can result from three fundamentally different situations: (i) the alignment of PDM vectors is completely random, (ii) all PDM vectors are lying in the substrate plane, or (iii) half of them is aligned upward and the other half downward, i.e.\ they are antiparallel aligned and cancel each other.
(And there is many more possibilities resulting in net $\Lambda = 0$.)

Especially this last scenario, i.e.\ mutual cancellation of PDMs is highly relevant in SOP films because, from basic electrostatics, antiparallel PDM alignment is energetically favored in the bulk of a system of polar molecules, which is considered to be the main cause of the relatively low $\Lambda$ values of neat films of common SOP materials, which are typically between about \SI{5}{} and \SI{10}{\%} only\cite{Noguchi2019,Hofmann2021}.
Thus, an important goal in controlling and engineering the SOP of organic semiconductor thin films is to elucidate the driving forces for PDM alignment at the surface of a film grown by PVD, and to tune them by suitable film preparation conditions.

\section{Results and Discussion}

\subsection{Controlling SOP by substrate temperature}

As pointed out previously\cite{Hofmann2021}, the so-called surface equilibration model has been established to enable control of optical anisotropy, i.e.\ birefringence or TDM alignment in absorption\cite{Lyubimov2015}, by the ratio of $T_\mathrm{S}/T_\mathrm{g}$.
In brief, the main idea is, that the motion of molecules on the surface of a film being grown in vacuum is orders of magnitude faster than it is in the bulk.
Thus, molecular orientation is determined at the surface of the film in the upper molecular layers, while it is largely maintained once the uppermost layer(s) are covered by newly deposited molecules.
For example, the surface equilibration model has been successfully applied to the optical orientation of \alq{}\cite{Bagchi2019}.
In that model, film growth shows a characteristic substrate temperature and rate dependence and a superposition of both is required to explain the findings, which has been shown by Bishop~et~al.\ by varying the evaporation rate by $\approx{}3$ orders of magnitude\cite{Bishop2019}.
In addition to that, it is known that the GSP of TPBi films grown at substrate temperatures higher than \SI{300}{\kelvin} decreases and even vanishes if $T_\mathrm{S}$ approaches the glass transition temperature of TPBi\cite{Bangsund2020}, where a so-called ultra-stable molecular glass is formed\cite{Rafols-Ribe2018}.
It was found to be beneficial for OLED efficiency because it avoids sub-turn-on exciton quenching by charge accumulations due to SOP in one part of the OLED.
However, it was also shown that charge accumulation can be reduced by engineering the SOP of the OLED as a whole\cite{Noguchi2022}.

\begin{figure}[tbp]
	\centering
	\includegraphics[width=\fscale\linewidth]{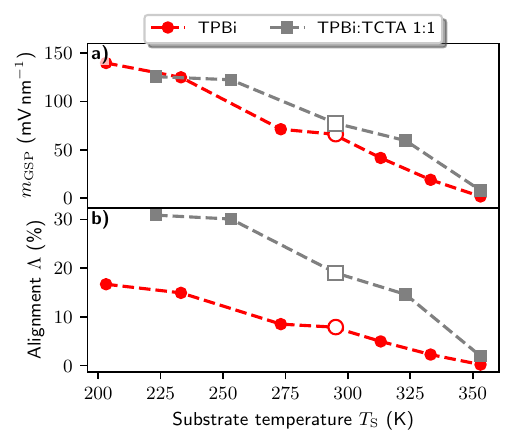}
	\caption{TPBi (circles) and TCTA:TPBi in 1:1 mixture (square) for different substrate temperatures. The GSP measured using Kelvin probe is given on the top (\textbf{a)}), the calculated alignment factor below (\textbf{b)}). Note that the GSP is similar for both, neat film TPBi and the mixture, requiring the molecular alignment to be roughly doubled in the latter. Room temperature is marked with a larger, open symbol.}
	\label{fig:tpbi-temp-mixing}
\end{figure}

In Figure~\ref{fig:tpbi-temp-mixing}, we show a variation of the substrate temperature $T_\mathrm{S}$ spanning roughly \SIrange{200}{350}{\kelvin} for both TPBi, and TPBi mixed with TCTA in a 1:1 ratio.
Although the mixture contains only \SI{50}{vol.\%} of the polar species, the measured GSP is almost equal to the neat TPBi film for each temperature, or even slightly higher.
Moreover, both follow the same dependence on substrate temperature, i.e. the GSP goes down above room temperature and vanishes at about $T_\mathrm{S}=\SI{350}{\kelvin}$, whereas it increases substantially for $T_\mathrm{S}<\SI{300}{\kelvin}$ and reaches more than twice the room temperature value at the lowest achievable $T_\mathrm{S}$.
This dependence is consistent with the surface equilibration model, which consequently seems to be valid for both neat film TPBi and the mixture.

From the model, we would expect orientation to start becoming isotropic for temperatures approaching the glass transition of the film at around $T_\mathrm{S} \approx 0.9T_\mathrm{g}$\cite{Dalal2015}.
In case of neat film TPBi ($T_\mathrm{g,TPBi} = \SI{123}{\celsius}$\cite{note_tgs}), this is expected to happen at around $T_\mathrm{S} = \SI{355}{\kelvin}$, which can be confirmed from Figure~\ref{fig:tpbi-temp-mixing}a.
Since TCTA has a higher $T_\mathrm{g,TCTA} = \SI{153}{\celsius}$\cite{Tsuchiya2022} than TPBi, we expect the mixture to have a slightly higher effective glass transition $T_\mathrm{g,mix}\approx\SI{137}{\celsius}$, as will be discussed in the next section in more detail.
Still, the value is close to where isotropic orientation can be inferred from the suppression of SOP seen in a vanishing GSP.
The trend is similar to what has been reported earlier for e.g.\ neat films of BCPO\cite{Cakaj2023}.

The similarity of their GSP directly implies that the oriented dipole fraction, i.e. the alignment parameter $\Lambda$ for the mixed film must be roughly twice the value of the neat TPBi film, which can be confirmed from Figure~\ref{fig:tpbi-temp-mixing}b.
For room temperature, for example, marked with an open symbol in Figure~\ref{fig:tpbi-temp-mixing}, we find $\Lambda\approx\SI{8}{\%}$ for neat TPBi, while it is $\approx\SI{20}{\%}$ in the TCTA:TPBi 1:1 mixture.
Please note, that this enhancement cannot be explained by the higher $T_\mathrm{g,mix}$ of the mixture, as will be shown later.
Further note that we kept the total rate of PVD of the two species the same as for the neat film growth so that we do not expect a major influence of the rate, as will be elaborated further below.

\subsection{Controlling SOP by mixing}

\begin{figure}[hb]
	\centering
	\includegraphics[width=\fscale\linewidth]{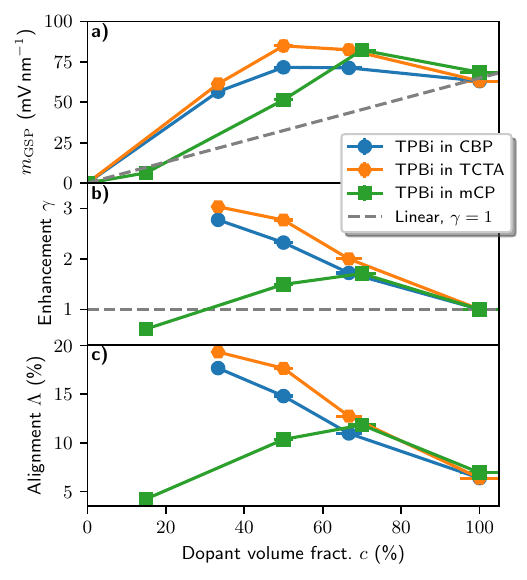}
	\caption{TPBi doped in CBP, TCTA and mCP. Data for CBP and TCTA has already been published in Reference~\citenum{Noguchi2020} and was re-evaluated for the purpose of this article. Shown are the GSP (\textbf{a)}), \textbf{b)} Enhancement factor $\gamma$ and \textbf{c)} the oriented dipole fraction $\Lambda$. In addition to the measurement, an exemplary non-enhancing, linear combination of TPBi with a non-polar host is drawn with a dashed line.}
	\label{fig:tpbi-mixing}
\end{figure}

Another approach to control SOP consists in mixing the polar species by co-evaporation with a non-polar one.
Through this "dilution" of the polar molecules, the resulting magnitude of PDM alignment can be enhanced.
This has first been reported by Jäger~et~al.\cite{Jaeger2016}, where an \alq{} layer diluted by \npb{} exhibits the highest interfacial charge density for a 1:1 mixture -- even higher than the neat-film \alq.
That said, molecular orientation is \textit{enhanced}, if the polar species shows higher alignment in diluted form than in neat film.
In measurement, such behavior will then lead to a non-linear shape of the GSP vs. concentration of the polar species, as seen in Figure~\ref{fig:tpbi-mixing}a.
Starting from \SI{100}{\%}, there is an increase in GSP for decreasing TPBi concentration in all mixtures.
However, with mCP as host, the maximum GSP is already achieved at \SI{75}{\%} TPBi, while CBP and TCTA have their maximum GSP at lower TPBi content of \SI{50}{\%}.

In order to normalize the GSP of the mixed films to the neat film values, an enhancement factor can be defined as follows for a polar species doped into a non-polar host\cite{Noguchi2020}:
\begin{equation}
\gamma = \frac{m_\mathrm{GSP,mix}\left(c\right)\,\varepsilon_\mathrm{mix}\left(c\right)}{c\cdot{}m_\mathrm{GSP,neat}\varepsilon_\mathrm{neat}}~,
\end{equation}
where $c$ is the concentration of the polar molecules, and $m_\mathrm{GSP,mix}\left(c\right)$ as well as  $\varepsilon_\mathrm{mix}\left(c\right)$ are the GSP and the dielectric constant of the mixture, respectively, while $m_\mathrm{GSP,neat}$ and $\varepsilon_\mathrm{neat}$ are GSP and dielectric constant of the neat film of the polar species, only.

Now one can clearly recognize a difference between the mCP host on the one hand side, where the enhancement reaches just a bit over 1.5 before dropping to significantly less than 1 at the lowest TPBi content, and the other two hosts (CBP and TCTA), where the enhancement factor keeps rising with decreasing TPBi concentration and reaches up to $3$ at $c=0.25$ (see Figure~\ref{fig:tpbi-mixing}b).
A similar behavior is observed for the SOP alignment parameter $\Lambda$ shown in Figure~\ref{fig:tpbi-mixing}c.
In CBP and TCTA the PDM alignment of the TPBi molecules increases from \SI{8}{\%} in the neat film to almost \SI{20}{\%} for a dilution $c=0.25$.
Such behavior has been reported before for TPBi and was ascribed to a reduction of electric dipole-dipole interactions between the PDMs of TPBi molecules\cite{Gunawardana2021}, if the material is mixed with a non-polar host.
However, with mCP as host we observe a different behavior; the maximum alignment is already achieved at $c=0.75$ before dropping to less than the neat film value at $c=0.15$.
Thus, in the latter host material we need to consider a different or an additional mechanism at play.

\subsection{Surface equilibration in mixed layers}

To compare the effect of substrate temperature variation (Fig.~\ref{fig:tpbi-temp-mixing}) with dilution (Fig.~\ref{fig:tpbi-mixing}), one has to be aware that mixing of two materials with different $T_\mathrm{g}$ modifies the \textit{effective} $T_\mathrm{g,mix}$ of the mixture.
The glass transition of a mixture is inferred from the Fox-equation\cite{Fox1956}
\begin{equation}
T_\mathrm{g,mix}(c) = \left(\frac{1-c}{T_\mathrm{host}} + \frac{c}{T_\mathrm{guest}}\right)^{-1}~,
\end{equation}
which reportedly provides a good estimate for the mixed film $T_\mathrm{g,mix}$ as long as the densities of the two species are the same (if not, the more general Gordon-Taylor equation has to be used)\cite{Tsuchiya2022}.
Please note, that the calculation is only valid if values are inserted in units of Kelvin.
The involved glass transitions range from $T_\mathrm{g,mCP} = \SI{64}{\celsius}$\cite{note_tgs} over $T_\mathrm{g,CBP} = \SI{110}{\celsius}$\cite{note_tgs} to $T_\mathrm{g,TCTA} = \SI{153}{\celsius}$\cite{note_tgs}, with TPBi centered in between ($T_\mathrm{g,TPBi} = \SI{123}{\celsius}$).

\begin{figure}
	\centering
	\includegraphics[width=\fscale\linewidth]{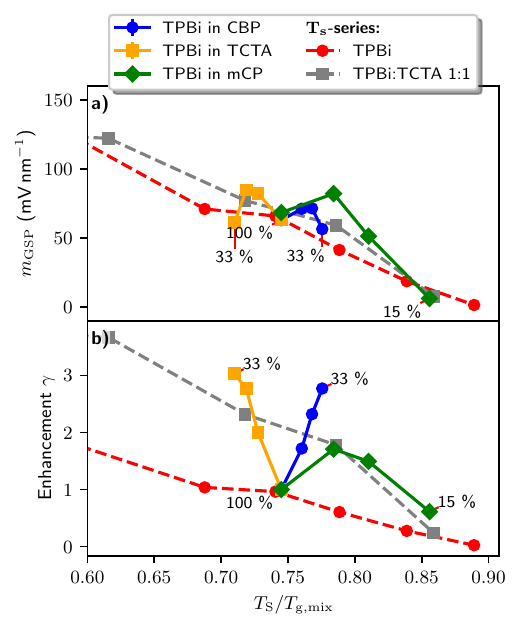}
	\caption{Three mixtures of TPBi in CBP, TCTA and mCP with \textit{constant} $T_\mathrm{S}$ but varied concentration alongside a variation of the substrate temperature (dashed lines) for neat TPBi and the TCTA:TPBi 1:1 mixture. Given is \textbf{a)} the measured GSP and \textbf{b)} the calculated enhancement factor $\gamma$ versus $T_\mathrm{S}/T_\mathrm{g,mix}$. Note that the comparison point with $T_\mathrm{S}=\SI{295}{\kelvin}$ located at $T_\mathrm{S}/T_\mathrm{g,mix} \approx 0.75$ is defined to be $\gamma = 1$.}
	\label{fig:tpbi-temp-more}
\end{figure}

Now, the three mixed films of TPBi diluted with TCTA, CBP and mCP can be compared with the neat TPBi film and the 1:1 mixture with TCTA grown at varied substrate temperature $T_\mathrm{S}$.
To this end, the GSP and enhancement data are plotted vs. $T_\mathrm{S}/T_\mathrm{g,(mix)}$, as shown in Figure~\ref{fig:tpbi-temp-more}.
While in neat TPBi as well as the 1:1 mixture with TCTA we varied $T_\mathrm{S}$, the spread over $T_\mathrm{S}/T_\mathrm{g,mix}$ for the mixtures is due to a change of $T_\mathrm{g,mix}$, only.
Looking at the raw GSP data (Figure~\ref{fig:tpbi-temp-more}a), they all seem to be close together and follow the already discussed trend described by the surface equilibration model: GSP increases for decreasing $T_\mathrm{S}/T_\mathrm{g,mix}$ and, vice versa, GSP vanishes for $T_\mathrm{S}/T_\mathrm{g,mix} \approx 0.85 - 0.9$.

However, the enhancement factor shown in Figure~\ref{fig:tpbi-temp-more}b tells a different story.
The behavior of the mCP:TPBi mixture still follows the trend of the TCTA:TPBi 1:1 mixture, indicating that, in this case, the effect of lowering the effective $T_\mathrm{g,mix}$ by adding the low-$T_\mathrm{g}$ host mCP to the neat TPBi is the dominant factor and, thus, obeys  surface equilibration.
But the other two mixing series with CBP and TCTA as host materials clearly have a much steeper dependence on $T_\mathrm{S}/T_\mathrm{g,mix}$; even more importantly, their slopes are opposite because CBP has a lower $T_\mathrm{g}$ than TPBi, while TCTA is higher.
This indicates that surface equilibration alone cannot describe their behavior.
Apparently, the achievable variation in the effective $T_\mathrm{S}/T_\mathrm{g,mix}$ is minor as compared to the effect of changing the dipole-dipole interactions of TPBi molecules upon dilution with a non-polar species.

To be precise, the glass transition temperature of the mixture alone is not the only parameter governing surface equilibration.
Recently, He~et~al.\ have used the rate-temperature superposition (RTS) model by Bishop~et~al.\cite{Bishop2019} to describe the interplay between growth rate and temperature in a more generalized way\cite{He2022}.
There, the effective growth rate is found to be exponentially dependent on $T_\mathrm{g}-T_\mathrm{S}$ with
\begin{equation}
R_\mathrm{eff} = R_\mathrm{act}\cdot{}10^{S\left(T_\mathrm{g}-T_\mathrm{S}\right)}~,
\end{equation}
where $S$, termed ``shift factor'' of unit \si{\dec\per\kelvin}, links rate and temperature together.
The authors determined $S\approx\SI{0.075}{dec\per\kelvin}$ for their data on neat TPBi and BPhen films (and 1:1 mixtures of both).
Interestingly, they were able to map all their data on one ``universal'' RTS curve, indicating that the behavior is fully described by surface equilibration.
As a result, a change of $T_\mathrm{S}$ by \SI{13.3}{\kelvin} equals a change in film growth rate by one order of magnitude.
Please note, that He~et~al.\ used a PDM magnitude of only \SI{2}{D} for TPBi\cite{He2022}, which explains their unusually high alignment value of about \SI{30}{\%} as an artefact of a too low PDM.
Another important point is that BPhen is polar as well, with a dipole moment of around \SI{3}{\debye} (BPhen is similar to BCP with a reported PDM of \SI{2.9}{\debye}\cite{Noguchi2019}).
Thus, it seems that dipole-dipole interactions in a BPhen:TPBi mixture are comparable to the neat TPBi film.
This is clearly not the case in our systems.

Nevertheless, the take-away message is still valid: a temperature change has a much greater effect on PDM orientation as changing the growth rate.
And, although a change in concentration implies a change in relative rate (within far below one order of magnitude) of the two species, the overall growth rate of the film can very well be kept constant.
Hence, a rate-effect can be ruled out in our case.

\subsection{SOP enhancement in other mixed systems}

As already mentioned, dipole-dipole interactions between the molecules' PDMs will act against molecular alignment and reduce the net film polarization.
Previously, the increasing distance between molecules has often been mentioned to qualitatively explain the enhanced alignment parameter upon dilution with a non-polar host\cite{Jaeger2016,Noguchi2020}.
To investigate the influence of a change in concentration $c$ on the magnitude of SOP, we can estimate the mean spacing of organic molecules in an amorphous thin film from its density, leading to a molecular spacing of $r \propto 1/\sqrt[3]{c}$.
The dipole-dipole interaction potential, in turn, is proportional to $1/r^3$ and, hence, for the net electrostatic energy of a diluted SOP system we find:
\begin{equation}\label{eq:dipolepotential}
V\left(c\right) \propto p^2n\cdot{}c~,
\end{equation}
where $n$ is the (number) density of the molecules and $p$ their dipole moment.
Note that, in this simple calculation, only one polar species is allowed, and the interaction between guest and host or induced dipoles is neglected.
As a result, the larger the concentration is, the stronger is the driving force to align PDMs in an anti-parallel manner, thus canceling SOP.
In particular, one may expect that the PDM magnitude of the polar species itself plays an important role for the dipole-dipole interactions, as will be discussed in the following.

\begin{figure}[tbp]
	\centering
	\includegraphics[width=\fscale\linewidth]{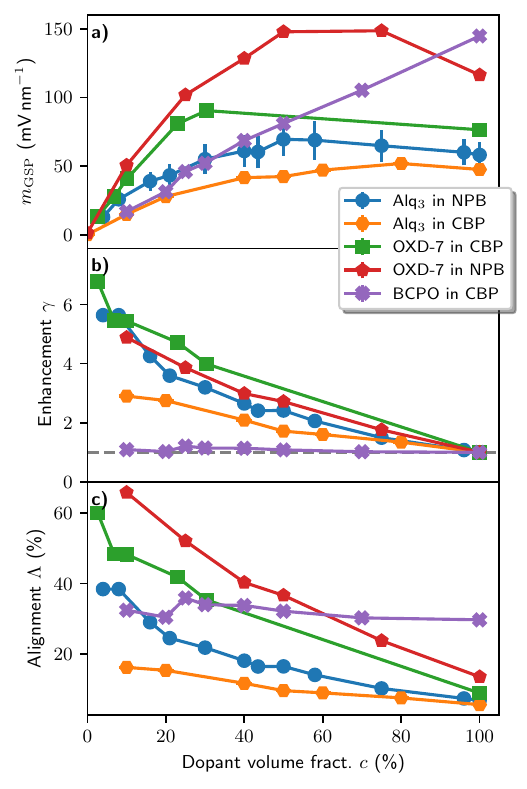}
	\caption{GSP, enhancement $\gamma$ and oriented dipole fraction $\Lambda$ of further  mixtures. Some of the GSP or interface charge data was already published elsewhere and was re-evaluated for this work using up-to-date values for material constants (see the supporting information for a complete list). Sources: NPB:\alq{} ref.~\citenum{Jaeger2016}, CBP:BCPO ref.~\citenum{Cakaj2023}, CBP:OXD-7 ref.~\citenum{Hofmann2020}.}
	\label{fig:further_enhancement}
\end{figure}

In Figure~\ref{fig:further_enhancement}, we show five additional measurement series with different polar species.
Apart from the GSP data, we also show the enhancement factor and the alignment parameter $\Lambda$ as well.
Already from the bare GSP data (Figure~\ref{fig:further_enhancement}a), one can distinguish two different types of behavior: While \alq{} and OXD-7 show a clearly non-linear dependence on concentration, similar to TPBi, BCPO is close to linear.
This becomes more evident when their respective enhancement factors and the degree of alignment are considered (see Fig.~\ref{fig:further_enhancement}b\&c).

Both, OXD-7 and \alq{} show considerable enhancement in CBP and NPB, with around 3 to 6 times more oriented dipoles for low concentrations, while the baseline oriented dipole fraction for neat films is below \SI{10}{\%}.
The dependence of $\gamma(c)$ on concentration is almost linear for all enhancing mixtures, just as previous observations with TPBi\cite{Noguchi2020}.
For OXD-7, three conformers with dipole moments between \SI{3.8}{D} and \SI{6.6}{D} exist, the weighted average is \SI{4.8}{D}\cite{Emelyanova2013}.
In case of \alq{}, only one isomer is expected in thin films\cite{Coelle2004}, with a dipole moment of around \SI{4.4}{D}\cite{Noguchi2013}.
Note, that for both OXD-7 and \alq{}, molecular dynamics simulations by Friederich~et~al.\ revealed a net PDM of \SI{6}{D}\cite{Friederich2018}, which would reduce the absolute values of $\Lambda$ to some extent, but not the overall trend, nor $\gamma$.

BPCO doped in CBP does not show enhancement, but a linearly rising GSP with $c_\mathrm{BCPO}$\cite{Cakaj2023}, where $p_\mathrm{BCPO}\approx\SI{3.5}{D}$\cite{Cakaj2023}.
The calculated enhancement factor is close to 1 for all concentrations, indicating no influence in the orienting force depending on the concentration of BCPO in CBP.
However, the overall orientation of BCPO in CBP is high, with around \SI{30}{\%} of dipoles contributing to the GSP already in the neat film.

Combining the results for both $\gamma(c)$ and $\Lambda(c)$, two findings are evident.
For one, in this small set of materials, $\gamma$ is larger, i.e.\ the enhancement is more pronounced, for molecules with a higher net PDM.
TPBi, however, discussed in the previous section, exhibits a dipole moment of almost \SI{6}{D} in thin film with its predominant C$_1$ isomer, and shows a maximum enhancement of only 3 to 4, see Figure~\ref{fig:tpbi-temp-more}.
Additionally, looking at $\Lambda$, those molecules with high enhancement (and large PDM) seem to show a lower oriented volume fraction in neat film -- and vice versa, those enhancing less, still show a high amount of overall orientation at $c = \SI{100}{\%}$.

To explain that dependence, we have to discuss the overall distribution of oriented dipoles in thin films in more detail.
Although a figure of merit like $\Lambda$ allows to specify the net alignment of PDMs, the same value might be achieved by totally different angular distributions, as $\Lambda$ only probes the $1^\mathrm{st}$ moment of the distribution function $f(\cos\theta_\mathrm{PDM})$.
However, we argue that an enhancement of SOP due to reduced dipole-dipole interaction is only expected in those cases, where a broader distribution of $\cos\theta_\mathrm{PDM}$ is observed, with a considerate amount of antiparallel orientation (see Reference~\citenum{Friederich2018} for a graphical depiction of $f$).
In other words: if the PDMs are anyway oriented mostly in a certain direction with a narrow distribution, no enhancement due to increased dipole-dipole distance is expected.

\begin{figure}
	\centering
	\includegraphics[width=\fscale\linewidth]{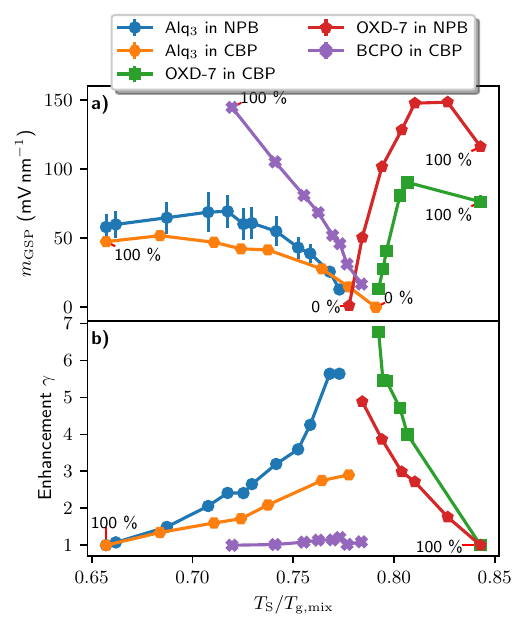}
	\caption{GSP and enhancement $\gamma$ series plotted versus $T_\mathrm{S}/T_\mathrm{g,mix}$ for the same datasets as in Figure~\ref{fig:further_enhancement}. Glass transition temperatures are again inferred from the Fox equation.}
	\label{fig:further_enhancement_tgs}
\end{figure}

For OXD-7, \alq{} and TPBi, the distribution function has been computationally investigated by Friederich~et~al., where OXD-7 and \alq{} exhibit a broad distribution of PDMs with a significant portion of all PDMs lying in plane, while TPBi seems to be oriented mostly up or down, with a tendency for the positive oriented PDM, thus leading to a GSP as well\cite{Friederich2018}.
In that work, they also estimate the importance of electrostatic interactions for the magnitude of the GSP.
Interestingly, for \alq{} and TPBi neat films they predict an increase of PDM alignment by almost a factor of 3, if electrostatic interaction is switched off in their simulation\cite{Friederich2018}.
However, for OXD-7, they predict a decrease of SOP, when electrostatics is switched off, contrary to the observation.
Figure~\ref{fig:further_enhancement_tgs} shows the series of \alq{}, OXD-7 and BCPO diluted in NPB and CBP plotted versus $T_\mathrm{S}/T_\mathrm{g,mix}$.
Here, one has to be aware, that the glass transition temperature of OXD-7 is relatively low ($T_\mathrm{g,OXD-7}\approx\SI{77}{\celsius}$\cite{note_tgs}).
Thus, mixing OXD-7 with CBP or NPB leads to an increase of $T_\mathrm{g,mix}$.
This means that part of the enhancement in diluted OXD-7 may also come from this effect.
Actually, the neat OXD-7 film has a $T_\mathrm{S}/T_\mathrm{g}$ close to $0.85$, i.e.\ close to the ultra-stable-glass region, where alignment vanishes.
Thus, its enhancement by mixing with CBP or NPB is overestimated, because it does not originate from PDM dilution alone, but also from lowering $T_\mathrm{S}/T_\mathrm{g,mix}$.
Or, in other words, for $c\left(\mathrm{OXD-7}\right) \gtrsim \SI{50}{\%}$ the GSP is being suppressed by the low glass transition, as is evident in Figure~\ref{fig:further_enhancement_tgs}a.

In case of BCPO, however, the direction of the PDM and the strongly orienting P=O group align, leading to a highly oriented neat film\cite{Cakaj2023}.
Thus, electrostatic dipole-dipole interactions do not seem to play an important role in this material and, hence, there is no enhancement upon PDM dilution.

This is similar to the case of CBP:\irppyacac{} (shown in the Supporting Information), where the chemical asymmetry of the molecule due to the (acac)-group leads to a high fraction of oriented dipoles for high concentrations of \irppyacac\cite{Jurow2016}, $\Lambda\left(\SI{100}{\%}\right)\approx\SI{25}{\%}$, where PDM of \irppyacac{} calculates to only \SI{2.0}{D} (Data published by Morgenstern~at~al.\cite{Morgenstern2018}, see Figure~S1 in the Supporting Information for details).
But still, upon dilution in CBP the degree of alignment can be improved.
For \irppyacac{}, we have given an estimate of a possible distribution function in Reference~\citenum{Morgenstern2018}, with a high probability of a preferred upright-standing orientation, which has been confirmed as well by simulations\cite{Friederich2017}.

\section{Conclusion}

Beside processing conditions, the mixing with a second component has been proven to be a powerful tool to manipulate and control SOP in organic thin films.
We have demonstrated, that the physics described by the surface equilibration resp.\ rate-temperature superposition models can indeed be applied to binary mixtures as well.
However, surface equilibration alone does not suffice to explain the huge enhancement of molecular orientation observed in some cases, but other effects like a suppression of dipole-dipole interaction by dilution have to be taken into account as well.
Overall, for each set of materials, multiple effects have to be considered, which can roughly be summarized as follows.

At first, molecules must have a tendency to \textbf{orient in a certain way}, wether in mixture or in neat film. Depending on the molecule, van der Waals interactions\cite{Morgenstern2018,Moon2017}, hydrogen bridges\cite{Cakaj2023} or a distinct shape anisotropy can favor orientation upon film growth\cite{Gunawardana2021}.

In contrast, molecules discussed in context of SOP are polar, with a sizable PDM.
The energetically favored orientation of two dipoles is, however, antiparallel -- which would effectively \textbf{suppress SOP}\cite{Gunawardana2021}.

At last, we find \textbf{mediating factors} not expected to lead to non-isotropic orientation in their own, but rather enhance or hinder existing interactions.
Most notable, the growth kinetics at the film surface, corned by the glass transition temperature and growth rate heavily influence molecular orientation\cite{Lyubimov2015,Bishop2019}.
Additionally, the mixing capabilities of two species can influence the resulting orientation.
Intermolecular interactions might cause two components not to mix well, which might increase the effect of suppressing factors\cite{Gunawardana2021,Noguchi2020}.

\section*{Acknowledgment}

We thank Deutsche Forschungsgemeinschaft (DFG) for funding of this work under grant agreements BR 1728/20-3 (project no.\ 341263954) and INST 94/163-1 (project no.\ 530008779), as well as Deutscher Akademischer Austauschdienst (DAAD) (project no.\ 57663382) for supporting exchange with Meiji University in Japan.

\providecommand{\latin}[1]{#1}
\makeatletter
\providecommand{\doi}
  {\begingroup\let\do\@makeother\dospecials
  \catcode`\{=1 \catcode`\}=2 \doi@aux}
\providecommand{\doi@aux}[1]{\endgroup\texttt{#1}}
\makeatother
\providecommand*\mcitethebibliography{\thebibliography}
\csname @ifundefined\endcsname{endmcitethebibliography}
  {\let\endmcitethebibliography\endthebibliography}{}

\end{document}